\documentclass[twocolumn,floatfix,showpacs,preprintnumbers,amsmath]{revtex4}

\usepackage{verbatim} 
\usepackage{amsmath} 
\usepackage{graphicx} 
\usepackage{color}
\usepackage{xcolor}
\usepackage{hyperref}

\usepackage[geometry]{ifsym}

\begin{document}

  \def\red#1{{\textcolor{red}{ #1}}}
  \def\blue#1{{\textcolor{blue}{ #1}}}
\title{
Questioning the relationship 
between the $\chi_4$ susceptibility and the dynamical correlation length 
in a glass former
}

\date{\today}

\author{R\'emy Colin~\footnote{R.C.'s current address:
Max Planck Institute for Terrestrial Microbiology,
Karl-von-Frisch Strasse 16, 35043 Marburg, Germany}\textit{$^{a}$}, 
Ahmed M. Alsayed~\textit{$^{b}$}, Cyprien Gay~\textit{$^{a}$} and
B\'ereng\`ere Abou~$^{\ast}$\textit{$^{a}$}}
\affiliation{%
\textit{$^{a}$} Laboratoire Mati\`ere et Syst\`emes Complexes, UMR CNRS 7057 \& Universit\'e Paris Diderot, 10 rue A. Domon et L. Duquet, 75205 Paris Cedex 13, France. \\
E-mail: remy.colin@univ-paris-diderot.fr; berengere.abou@univ-paris-diderot.fr; \\
\textit{$^{b}$} Complex Assemblies of Soft Matter Laboratory, UMI CNRS 3254, Rhodia INC., 350 G. Patterson Blvd, Bristol PA 19007, USA.}

\newcommand{\dg}{\,^\circ}
\newcommand{\T}{\mathrm{T}}
\newcommand{ \dd}{\mathrm{d}}
\newcommand{\pixel}{\mathrm{px}}
\newcommand{\MSD}{\langle \Delta r^2\left(\tau\right)\rangle}
\newcommand{\fps}{\mathrm{im./s}}
\newcommand{\der}[2]{\frac{\mathrm{d} #1}{\mathrm{d} #2}}
\newcommand{\Der}[2]{\frac{\mathrm{d} }{\mathrm{d} #2}#1}
\newcommand{\dervar}[2]{\frac{\partial #1}{\partial #2}}
\newcommand{\G}{\mathrm{G}}
\newcommand{\MSDi}{\langle \Delta r_i^2\left(t\right)\rangle_{t'}}
\newcommand{\Neff}{N_{\mathit{eff}}}
\newcommand{\micron}{\mathrm{\mu m}}
\newcommand{\millimeter}{\mathrm{mm}}
\newcommand{\pxl}{l} 

\newcommand{\ur}{{\boldsymbol{\hat{u}_r}}}
\newcommand{\tenso}[1]{\underline{\underline{\tens{#1}}}}
\newcommand{\utheta}{\boldsymbol{\hat{u}_\theta}}
\newcommand{\uphi}{\boldsymbol{\hat{u}_\varphi}}
\newcommand{\mfric}{{\zeta}_R}
\newcommand{\omeg}{\boldsymbol{{\omega}^\bot}}
\newcommand{\fric}{{\zeta_R}^\bot}
\newcommand{\MSAD}{\langle \Delta \psi^2(t) \rangle}
\newcommand{\Dr}{D_{R}}
\newcommand{\Lm}{L}
\newcommand{\MSDphi}{\langle \Delta r_{\varphi}^2(t) \rangle}
\newcommand{\MSDtheta}{\langle \Delta r_{\theta}^2(t) \rangle}
\newcommand{\MSDr}{\langle \Delta r_{r}^2(t) \rangle}

\newcommand{\R}{\vect{\mathcal{R}}}
\newcommand{\cotan}{\mathrm{cotan}}
\newcommand{\Du}{\Delta \ur}
\newcommand{\vect}[1]{\mathbf{#1}}

\newcommand{\vkp}{\vect{k^\prime}}
\newcommand{\vk}{\vect{k}}
\newcommand{\vrr}{\vect{r}}
\newcommand{\vp}{\vect{p}}
\newcommand{\taur}{\tau_{\rm r}}
\newcommand{\Qtau}{Q(\tau)}
\newcommand{\Qt}{Q_{t}(\tau)}
\newcommand{\qk}{q_{\vect{k}}^{\tau}}
\newcommand{\qkp}{q_{\vect{k^\prime}}^{\tau}}
\newcommand{\qkr}{q_{\vect{k+r}}^{\tau}}
\newcommand{\qkt}{q_{\vect{k},t}^{\tau}}
\newcommand{\qkpt}{q_{\vect{k}^\prime,t}^{\tau}}
\newcommand{\qkrt}{q_{\vect{k+r},t}^{\tau}}
\newcommand{\la}{\left\langle}
\newcommand{\ra}{\right\rangle}
\newcommand{\roi}{ROI}
\newcommand{\rd}[1]{\textcolor{red}{#1}}
\newcommand{\g}[1]{\textcolor{green}{#1}}
\newcommand{\f}[1]{\textcolor{magenta}{#1}}

\begin{abstract}
Clusters of fast and slow correlated particles, identified as dynamical heterogeneities (DHs),
constitute a central aspect of glassy dynamics.
A key ingredient of the glass transition scenario
is a significant increase of the cluster size $\xi_4$ as the transition is approached.
In need of easy-to-compute tools to measure $\xi_4$, the dynamical susceptibility $\chi_4$ 
was introduced recently, and used in various experimental works to probe DHs.
Here, we investigate DHs in dense microgel suspensions using image correlation analysis,
and compute both $\chi_4$ and the four-point correlation function $G_4$. 
The spatial decrease of $G_4$ provides a direct access to $\xi_4$, 
which is found to grow significantly with increasing volume fraction.
However, this increase is not captured by $\chi_4$.
We show that the assumptions that validate the connection between $\chi_4$ and $\xi_4$
are not fulfilled in our experiments.\newline
{\bf The present version was accepted for publication in \href{http://pubs.rsc.org/en/journals/journalissues/sm}{\em Soft Matter}}
\end{abstract}


\maketitle

\section*{Introduction}

Understanding the glass transition and
the out-of-equilibrium glassy dynamics remains a challenge in condensed matter
physics. In practice, glass transitions are observed in various systems,
such as molecular liquids, colloids or granular
materials \cite{Science-Angell-1995, berthier_theoretical_2011,
  liu_jamming_2010}. Among all, dense colloids are
model systems with a glass transition at ambient temperature upon increasing volume
fraction \cite{nature-Pusey-1986}. They display slow but accessible
timescales and can be probed with simple optical techniques such as
microscopy and dynamic light scattering \cite{Science-Weeks-2000, Science-Kegel-2000}.

Over the last $15$ years, dynamical heterogeneities (DHs) have been
recognized as a promising feature in understanding slow relaxation
processes in glass-forming systems \cite{AnnRevGlass-Ediger-2000,
  JPCM-Richert-2002, JNCS-Sillescu-1999}. 
  DHs consist of fast and slow clusters of dynamically correlated particles, coexisting
in the material, with the idea that a dynamical correlation length --
representing the clusters size -- diverges when approaching the
glass transition \citep{PRE-Marcus-1999,JCP-Glotzer-2003,
  nphys-Wang-2006,PRL-Chaudhuri-2007,Colin-DH-2010,biroli_random_2009,
  Science-Hedges-2009}. Dynamical heterogeneities are predicted by
theories and have been observed in numerical simulations
\citep{PRE-Berthier-2004,JCP-Glotzer-2003, heterogBerthier2,
  dasgupta_is_1991, donati_theory_2002, PRE-Vogel-2004,
  chandler_lengthscale_2006, parsaeian_growth_2008} and experimental
works \citep{Science-Berthier-2005, Nordstrom-DH-2011,
  nphys-Ballesta-2008, PRE-Abate-2007, epl-Cipelletti-2006,
  Sessoms-2009, JPCM-Weeks-2007, narumi_SM_2011,
  epl-Lechenault-2008-2}. 
They can mainly be quantified with
tools such as four-point correlation functions $G_4$, whose spatial
dependence gives a direct access to the dynamical correlation length
$\xi_4$, or with 
dynamical susceptibilities $\chi_4$ which
have recently been proposed as easy-to-compute indirect
tools~\cite{JCP-Glotzer-2003,Science-Berthier-2005}.

Here, we investigate DHs with both a four-point correlation function
$G_4$ and its associated dynamical susceptiblity
$\chi_4$, in dense suspensions of soft microgel particles, by performing image correlation analysis. 
With the direct tool $G_4$, we measure a significant growth of
the dynamical correlation length $\xi_4$ with increasing volume
fraction. We then investigate the validity of the dynamical susceptibility $\chi_4$
as a tool to extract the dynamical correlation length $\xi_4$, 
and analyze the reasons why it fails to quantify the growth of $\xi_4$.

\section{Theoretical tools for measuring the correlation length}

We consider a system described by a local order parameter
$q_{\vect{j},t}(\tau)$, here defined as the time correlation 
of the observable quantity ({\it e.g.} local density, particle position, 
transmitted light intensity) between time $t$ and $t+\tau$ at point $\vect{j}$.
A space-averaged 
and a space-time-averaged order parameters, 
$Q_t(\tau)=\langle q_{\vect{j},t}(\tau)\rangle_{\vect{j}}$
and
$Q(\tau)=\langle Q_t(\tau)\rangle_{t}$ respectively, are
constructed so that the averaged time-correlation function 
$Q(\tau)$
measures the relaxation dynamics of the system. 
This quantity decays from $1$ to $0$ as particles move a
characteristic distance between $t$ and $t+\tau$
\cite{heterogDauchot2}. 

A direct route to the correlation length $\xi_4$ is the
4-point correlator $G_4^{\rm vect}(\vect{r},\tau)$ 
which reflects how $q_{\vect{j},t}(\tau)$ is correlated between
points separated by $\vrr$:
\begin{multline}
G_{4}^{\rm vect}(\vect{r},\tau) = 
 \langle (q_{\vect{j}+\vect{r},t}(\tau) - q_{\vect{j}+\vect{r}}(\tau)) 
(q_{\vect{j},t}(\tau)-q_{\vect{j}}(\tau)) \rangle_{\vect{j},t}  \\ = 
\langle  q_{\vect{j}+\vect{r},t}(\tau)q_{\vect{j},t}(\tau) \rangle_{\vect{j},t} 
- \langle q_{\vect{j}+\vect{r}}(\tau) q_{\vect{j}} (\tau) \rangle_{\vect{j}} 
\label{G4-def-nonorm}
\end{multline} 
where $q_{\vect{j}}(\tau) = \langle q_{\vect{j},t}(\tau)\rangle_t$. 
The corresponding 4-point correlation function $G_4(r,\tau)$ is
defined as, $G_4(r,\tau)=\langle G_4^{\rm
  vect}(\vect{r},\tau)\rangle_{r<|\vect{r}|<r+\delta r}$, with $\delta
r$ chosen such that the average runs over a sufficient number of
points.
%
In a system with a dominant dynamical correlation length scale $\xi_4(\tau)$,
the correlation function decays at large $r$ as: 
\begin{equation}
\label{eq:G4:Atau:exp:p}
G_{4}(r,\tau)
\sim \frac{1}{r^p} \exp(-r/\xi_4(\tau))
\end{equation}
where exponent $p$ is discussed below.
Equation~\eqref{eq:G4:Atau:exp:p} defines 
the dynamical correlation length $\xi_4$.

Another tool was introduced recently to characterize 
dynamical heterogeneities, namely the dynamical susceptibility $\chi_4(\tau)$
(see \cite{heterogDauchot2,heterogBerthier2} for a review).
It can be measured as the variance 
of the temporal fluctuations of $Q_t(\tau)$:
\begin{equation}
\chi_4(\tau) = N (\langle
Q^2_t(\tau)\rangle_{t} - \langle Q_t(\tau) \rangle^2_{t})
\label{def-chi4}
\end{equation}
with $N$ the number of points in space under consideration. 
Let us see how $\chi_4$
is indirectly connected to $\xi_4$. Since $Q_t(\tau)\equiv\la q_{\vect{j},t}(\tau)\ra_{\vect{j}}$, 
it can be shown that
$\chi_4$ is related to $G_4$ by $\chi_4(\tau)=\sum_{\vrr} G_4^{\rm vect}(\vrr,\tau)$
which can be expressed in the continuous limit as :
\begin{equation}
\chi_4(\tau)=\rho \int{\rm d}^2\vrr \, G_4(|\vrr|,\tau)
\label{interpr-chi4}
\end{equation}
with $\rho$ the average density of points in space.
It was recently proposed 
to use
$G_{4}(r,\tau) \sim A(\tau)\,/r^p \exp(-r/\xi_4(\tau))$
and Eq.~\eqref{interpr-chi4}
to clarify the link between $\chi_4(\tau)$ and $\xi_4(\tau)$.
In two dimensions, $\chi_4(\tau)$ is then :
\begin{equation}
\chi_4(\tau) \sim A(\tau)\, 2 \pi \rho \,\xi_4^{2-p}(\tau)
\label{chi4-sigma2B}
\end{equation}
The peak of $\chi_4(\tau)$ was proposed and used widely in numerical simulations and
experiments \cite{PRE-Abate-2007, crauste-thibierge_evidence_2010,
  PRE-Elmasri-2010, Science-Berthier-2005,Nordstrom-DH-2011}
to determine the time for which the dynamics is the most heterogeneous 
and indirectly, the correlation length value $\xi_4$. 
Here, the exponent $p$ can be related to the clusters fractal dimension or, 
if the clusters are all compact, to the cluster size distribution. 
As stated in \cite{heterogBerthier2}, the growing peak in $\chi_4$ 
upon increasing volume fraction reveals the growth of a dynamical correlation length 
if the assumptions made for the scaling of $G_4$ are
fulfilled.

\section{Materials and Methods}
%
\begin{figure}
\includegraphics[width=\columnwidth,clip]{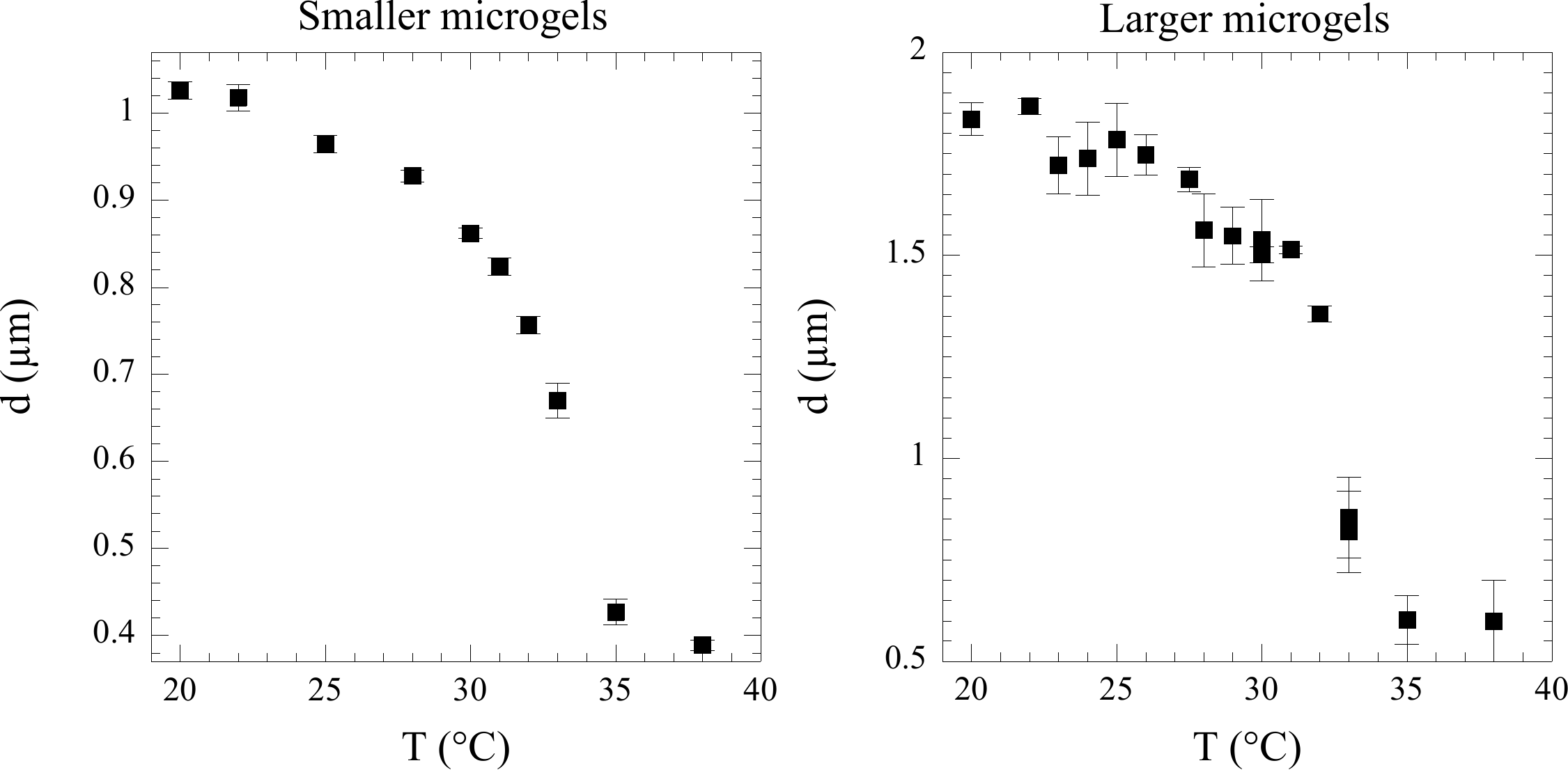} 
\caption{Microgels diameter as a function of temperature, measured with Dynamic Light Scattering. In order to suppress crystallisation, our suspension is prepared as a bidisperse
mixture of
small microgels (left) and large ones (right) 
(the number fraction of large particles is $18 \%$). The microgels
diameter decreases over the investigated temperature range (20-30$\dg$C),
with a constant diameter ratio 1:1.8. 
}
\label{diameter}
\end{figure}
\begin{figure}
\includegraphics[width=\columnwidth,clip]{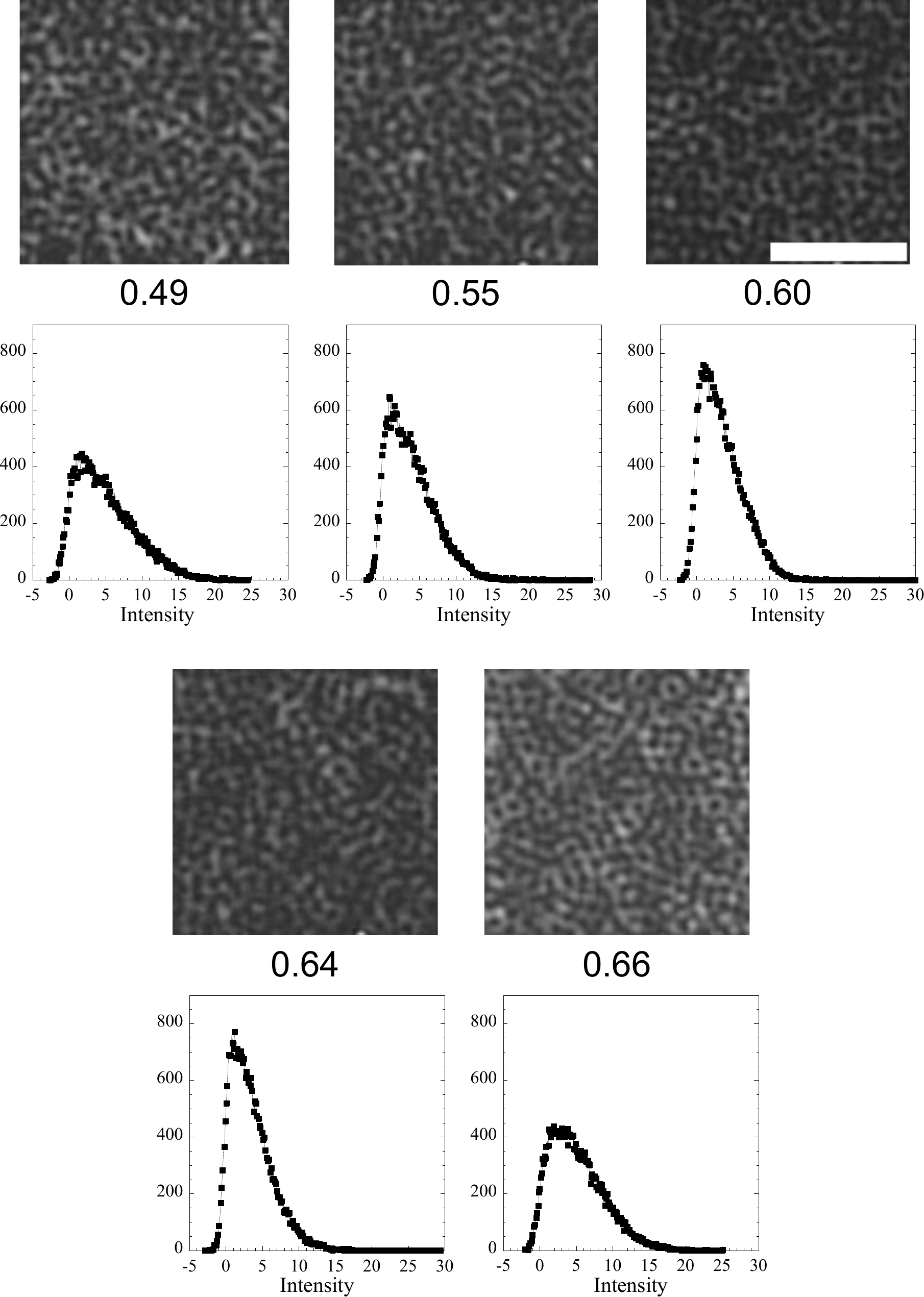}
\caption{Images of the microgel suspension at various volume fractions, ($0.49$,$0.55$,$0.60$, $0.64$ 
and $0.66$) in bright field microscopy at $\times 100$ magnification (oil immersion objective, NA=1.3); 
the subsets dimension is $187 \times 187$ pixel$^2$. The white bar is $96$ pixel long ($7\,\mu$m). 
Images have been processed 
(see details in the appendix)
to filter out the noise before image correlation analysis. 
The histogram of pixel intensities of these processed images is shown. The condenser aperture stop was set to the same setting for each experiment (condenser NA=0.55).
}
\label{image}
\end{figure}
{\bf Microgel suspension preparation.} Our model glass consists of a suspension 
of thermosensitive microgels, made of pNIPAm (poly-N-isopropylacrylamide) crosslinked
with BIS (N,N'-methylenebisacrylamide), whose synthesis is described
in \cite{Colin-DH-2010}. The microgel diameter decreases with
temperature, which provides a unique way of tuning volume fraction
with an external parameter~\citep{ACIS-Saunders-1999,
  microgel-suspensions_2011}. The suspension was
prepared by mixing small and large microgels with a
diameter ratio 1:1.8, constant over the investigated temperature range $20-30\dg$C (number fraction of large particles : $18 \%$). The bidispersity was used to suppress crystallisation.
Figure \ref{diameter} shows the microgel diameters as a function of temperature, 
measured with Dynamic Light Scattering. The diameters decrease by $\sim 20\%$ over the investigated temperature range. Within the $20-30\dg$C range, the pNIPAm particles behave as hydrophilic repulsive soft spheres 
(the transition to hydrophobic attractive spheres occurs at $32-33\dg$C 
and is easily identified by microscopy).

{\bf Effective volume fraction of the suspension}. At temperature $\T=30\dg$C, 
an effective volume fraction was assigned to the microgel suspension. 
For this purpose, latex probes (radius $R=0.5\,\mu$m comparable to the microgel radius) 
were added to the suspension and their mean-squared displacement (MSD) measured. 
The MSD was found to increase linearly with time, 
which defines a diffusion coefficient $D$, $\MSD_{i,t} = 4 D \tau$. 
A suspension viscosity $\eta=45$ mPa.s was deduced from the Stokes-Einstein relation, 
$\eta = k_B \T / 6 \pi R D$, expected to be valid at such low viscosity. 
The main issue with using the Stokes- Einstein relation in this case 
is that the size of the probe is similar to the size of the pNIPAm particle. 
This issue is tackled by considering the long time diffusion coefficient 
which shows the response to the medium on larger length scales. 
Following previous works performed in pNIPAm suspensions similar to ours~\cite{JCP-Senff-1999}, 
the suspension volume fraction was estimated from the measured value $\eta/\eta_{\rm solvent}$, 
using the empirical expression given in this paper. 
In our case, the volume fraction at $T=30\dg$C was approximately $\Phi_{30}=0.49$. 
The volume fractions at lower temperatures $T$ are then calculated 
using the relation 
$\Phi_{\rm eff}(\T) = (\bar{d}(\T) / \bar{d}(30\dg{\rm C}))^3 \, \Phi_{30}$, 
where $\bar{d}(T)$ is the number-averaged diameter $0.82\,d_{\rm small}+0.18\,d_{\rm large}$ 
at temperature $T$.

{\bf Sample preparation and video recording}. The microgel suspension was injected in a $3\times 3 \,{\millimeter}^2$
chamber made of a microscope plate and
a coverslip separated by a $250\,\micron$
thick adhesive spacer. The chamber was sealed with araldite glue to avoid evaporation and contamination. 
The samples were observed using standard bright field microscopy 
on a inverted Leica DM IRB microscope 
at $\times 100$ magnification (oil immersion objective, NA=1.3, depth of focus: $\sim 200$ nm). 
Typical images of the suspension at various volume fractions can be seen in Figure \ref{image}. 
The objective temperature was adjusted with a Bioptechs objective heater within $\pm 0.1 \dg$C. 
The sample temperature was maintained through the immersion oil in contact. 
A CCD camera (FOculus 124B) coupled to the microscope, was recording films of the microgel suspension. 
The camera was running at a frame rate from $30$ down to $0.375$ fps 
for a few minutes to several hours,
depending on the suspension dynamics. The region of observation was chosen at least
$100\,\micron$ away from the sample edges to avoid boundary
effects.

{\bf Image correlation analysis.} Films of the microgel suspension 
were analyzed with image correlation analysis, 
a suitable technique when particle trajectories cannot be resolved individually. 
Films consist of successive video frames ($640 \times 480$ pixel$^2$). 
Frames are first pre-processed following a procedure described in appendix. 
Image correlation analysis was then performed on $187 \times 187$ pixel$^2$ subsets 
of the pre-processed frames using home-made ImageJ~\cite{ImageJ} plugin 
(subset dimension: $13.6\,\times\,13.6\,\mu{\rm m}^2$, $1$ px $= 72.75$ nm). 
Each frame subset is described by a matrix of pixels $\vect{p}$ of intensity
$I_{\vect{p}}(t)$, where $\vect{p}$ is the two-dimensional index of
the pixel position and $t$ the time position of the frame in the
film. 
Possible global variations of
illumination and contrast along the film
were wiped out using
the normalised frame intensity $i_{\vect{p}}(t)$ at pixel
$\vect{p}$:
\begin{equation}
i_{\vect{p}}(t)=\frac{I_\vect{p}(t) - \langle I_\vect{p}(t)
\rangle_\vect{p}}{\sqrt{\langle (\delta I_\vect{p}(t))^2 \rangle_\vect{p}}}
\end{equation}
with $ \delta I_\vect{p}(t)= I_\vect{p}(t) -\langle I_\vect{p}(t)\rangle_\vect{p}$, 
and $\langle\cdot\rangle_\vect{p}$ the average over all the pixels in the
frame. 
The subsets are divided into
 non-overlapping squared Regions Of Interest, ${\rm ROI}$,
of dimension equivalent to a particle size, $11 \times 11$ pixel$^2$ 
($289$ ${\rm ROIs}$ in a subset). The subset
 intensity $i_{\vect{p}}(t)$ is now denoted as
 $i_{\vect{j,j'}}(t) \equiv i_{\vect{p}}(t)$, with $\vect{j}$ the
 center of ${\rm ROI}[\vect{j}]$ 
and $\vect{j'} \in {\rm ROI}[\vect{j}]$. (${\rm ROI}[\vect{j}]$ is 
centered around pixel
 $\vect{j}$. )
For each subset, following~\cite{duri_resolving_2009}, a local order
parameter, $q_{\vect{j},t}(\tau)$, was defined as,
\begin{equation}
q_{\vect{j},t}(\tau) = \langle i_\vect{j,j'}(t+\tau) i_\vect{j,j'}(t)  \rangle_{\vect{j'}\in ROI[\vect{j}]},
\label{defQit}
\end{equation}
The functions $Q(\tau)=\la q_{\vect{j},t}(\tau)\ra_{\vect{j},t}$, $\chi_4(\tau)$ and $G_4(r,\tau)$ 
were then derived from $q_{\vect{j},t}(\tau)$ using our home-made ImageJ plugin.

\section{Dynamical susceptibility $\chi_4$}
%
\begin{figure}
\includegraphics[width=0.65\columnwidth,clip]{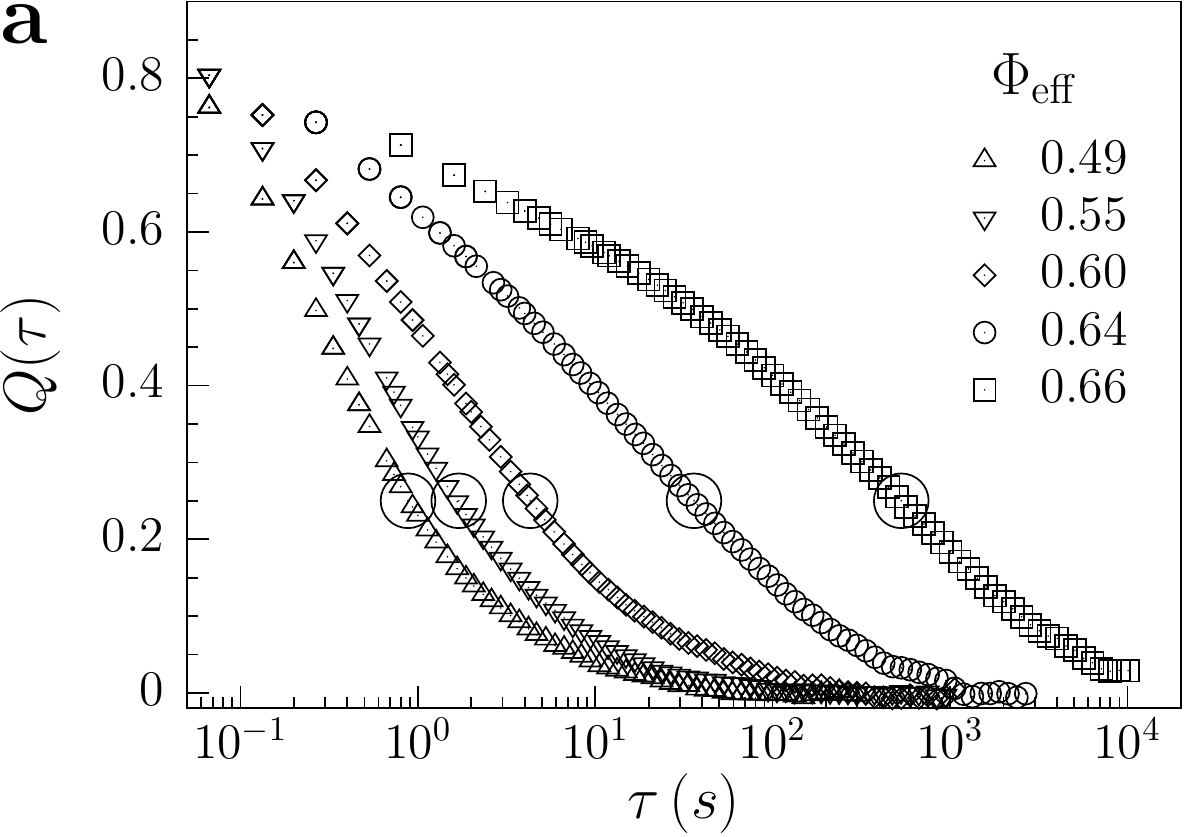}
\includegraphics[width=0.65\columnwidth,clip]{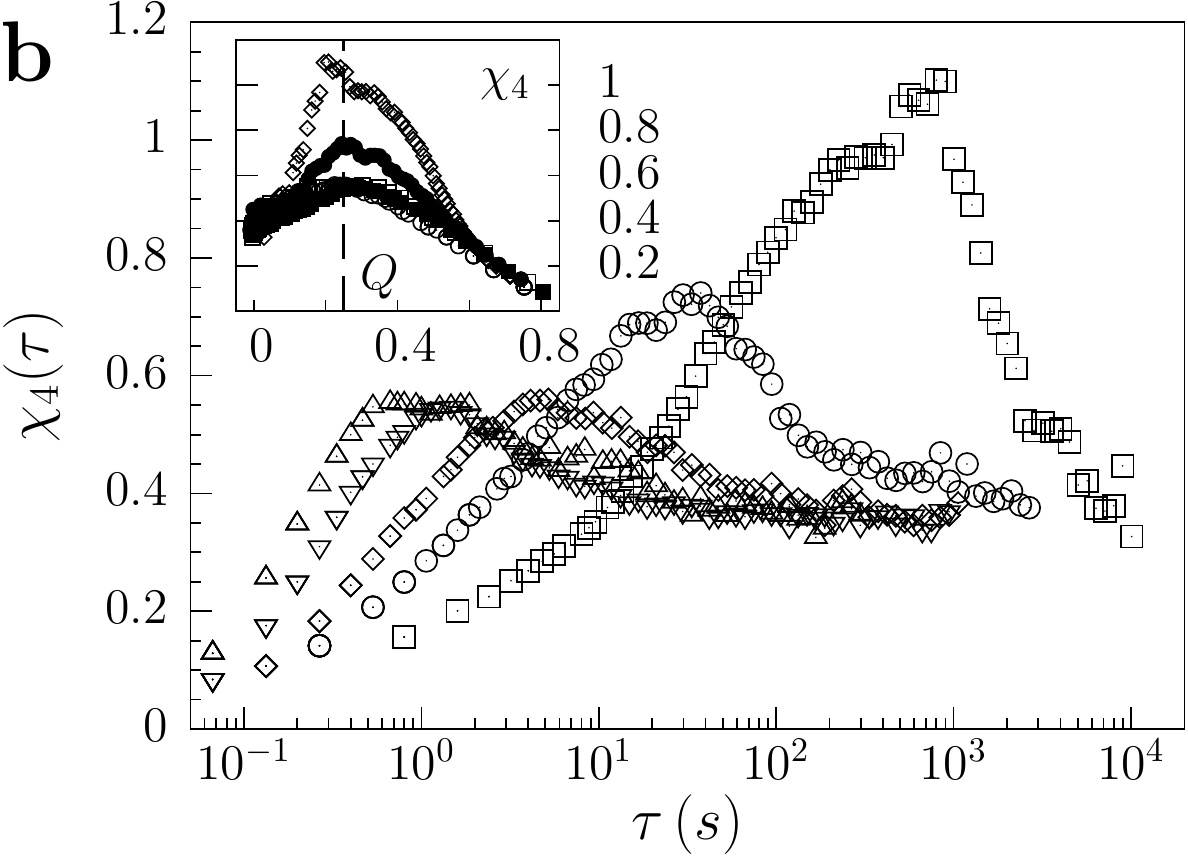}
\caption{Mean correlation function $Q(\tau)$ and dynamical
    susceptibility $\chi_4(\tau)$ measured in the microgel
  suspension with increasing volume fraction. In (a), 
the values of
the typical decay
  time 
$\tau^\star$
of $Q(\tau)$, 
denoted by large empty circles,
are: 
$0.88$, $1.7$, $4.3$, $36$ and $5.3\,10^2\,{\rm s}$.
In (b), the peak value of $\chi_4$ increases with volume fraction, suggesting
  the increase of a spatial correlation length. 
The inset reveals that the peak value is reached
approximately for $Q(\tau)=0.25$ (vertical line), 
the value chosen to estimate the typical decay time $\tau^\star$.
}
\label{functionQ}
\end{figure}
The microgel suspension was investigated
at various volume fractions in the {\em supercooled states}, 
at volume fractions below the glass transition~\cite{Colin-DH-2010}. 
The volume fraction was increased in a
quasistatic way, allowing the system to relax between each step and
reach an equilibrium state. Figure \ref{functionQ}-a 
shows the ensemble-averaged order parameter $Q(\tau)=\la q_{\vect{j},t}(\tau)\ra_{\vect{j},t}$, 
similar to the one in \cite{JPCM-Cipelletti-2003,Nordstrom-DH-2011,duri_resolving_2009}
in the microgel suspension at various volume fractions. 
It
decreases from its maximal value -- ideally $1$ at $\tau=0$ -- to $0$
at the largest lag times, with a typical decay time $\tau^\star$. It is
intuitive that $\tau^\star$ is related to the particles dynamics~: as the
particles get farther from their initial positions with increasing
lag time $\tau$, the time correlation function $Q(\tau)$ between two frames 
separated by $\tau$ gets smaller on average. With increasing volume fraction, 
the relaxation time $\tau^\star$, 
defined by $Q(\tau^\star) =0.25$~\footnote{The decay time $\tau^\star$ 
can be arbitrarily defined as $Q(\tau^\star) =0.25$. 
In the range $0.2 < Q(\tau) < 0.4$, the
  curves of $Q$ are parallel to each other and the increase in $\tau^\star$ 
  with volume fraction does not depend much on the value of
  $Q$.},
increases by three orders of magnitude, revealing the suspension dynamics slowing down.

The dynamical susceptibility $\chi_4(\tau)$ is shown in Figure
\ref{functionQ}-b. It exhibits a maximum whose value increases
with volume fraction. In the inset, $\chi_4(\tau)$ is plotted as a
function of $Q(\tau)$.
Its maximum is reached for
approximately the same value of $Q(\tau)$, corresponding to the suspension relaxation time
$\tau^\star$. 
Under the assumptions detailed in \cite{heterogBerthier2},
the increase of the peak value $\chi_4(\tau^\star)$ with volume fraction suggests
an increase of the spatial correlation length $\xi_4(\tau^\star)$ 
with volume fraction, see Eq.~\eqref{chi4-sigma2B}. 
In the following, we measure $\xi_4(\tau^\star)$ from $G_4(r,\tau^\star)$
and test the reliability of Eq.~\eqref{chi4-sigma2B}.

\section{Direct measurement of the spatial correlation length $\xi_4$}

\begin{figure}
\includegraphics[width=0.65\columnwidth,clip]{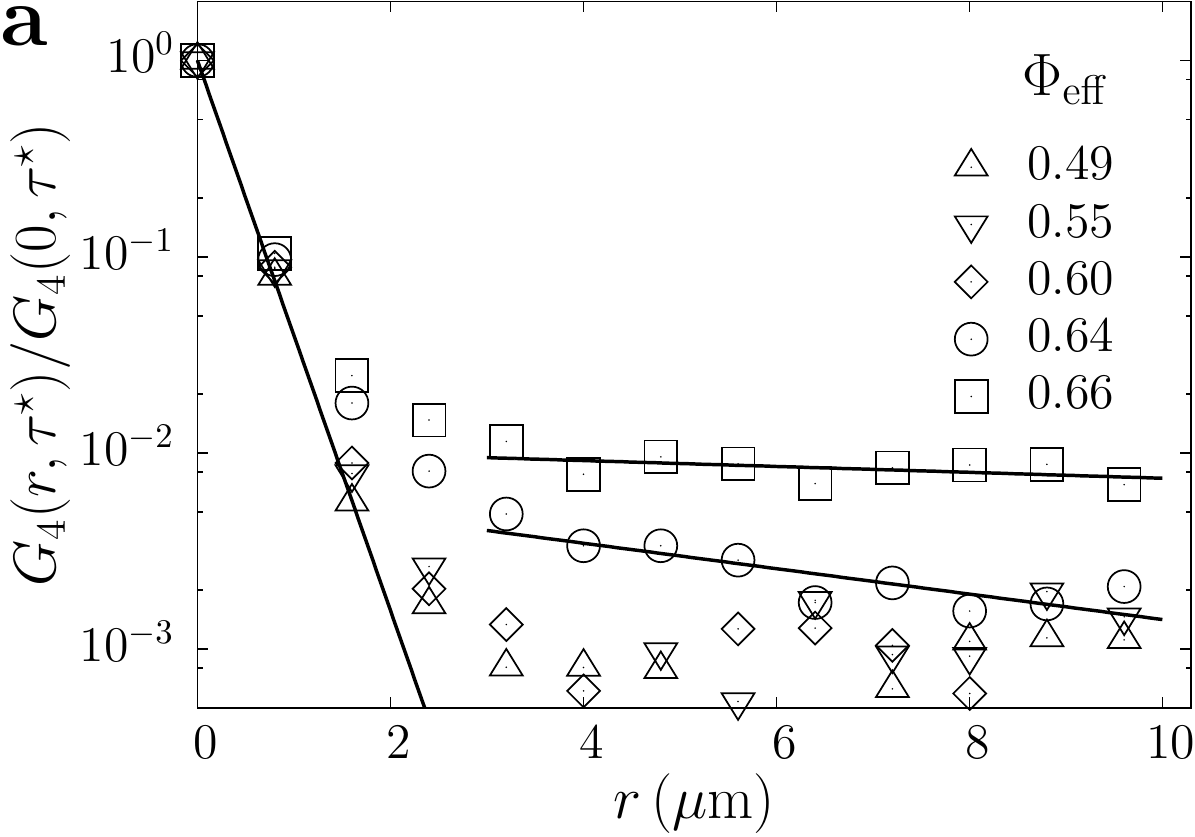}
\includegraphics[width=0.65\columnwidth,clip]{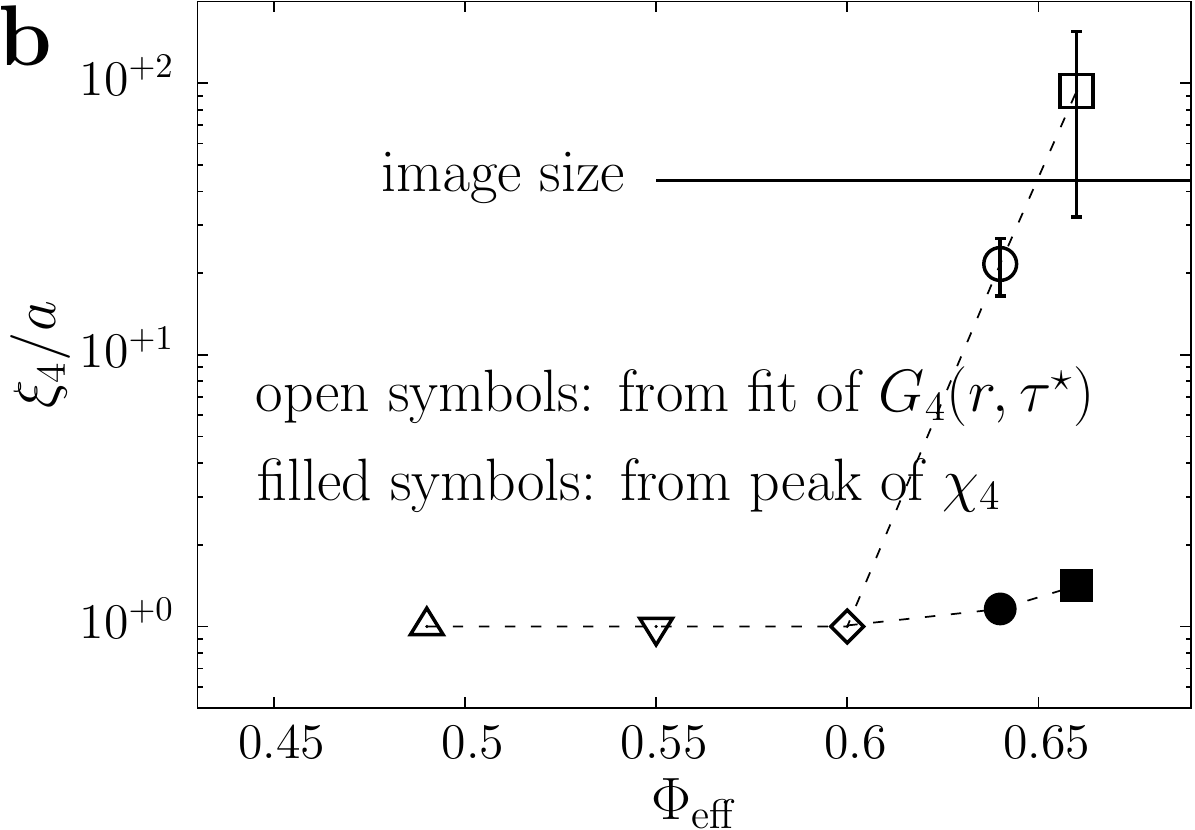}
\caption{Normalised 4-point correlation function $G_4(r,\tau^\star)/G_4(0,\tau^\star)$ 
and correlation length $\xi_4(\tau^\star)$. (a): 
We define the critical radius $r_c$ as
  the intersection of both asymptotes.
The solid lines correspond to fits according to Eq.~\eqref{functional1} with $p=0$.
(b): The correlation length $\xi_4$ extracted from the fits
of $G_4$ (open symbols) increases significantly with volume fraction.
The highest value of $\xi_4$ is larger than the image size (horizontal line)
and must be considered with caution.
  Values: 
  $\xi_4/a = 93 \pm 63$   at $\Phi_{\rm eff}=0.66$;
  $\xi_4/a = 22 \pm 5$ at $\Phi_{\rm eff}=0.64$; 
At $\Phi_{\rm eff}=0.60$, $0.55$ and $0.49$,
the second decay cannot be quantified
and we identify the correlation length with the effective particle radius, $\xi_4=a$. 
The values of the correlation length 
derived from the peak value of $\chi_4$ 
(filled symbols)
are significantly lower.
}
\label{g4_mes}
\end{figure}
\begin{figure}
\includegraphics[width=0.66\columnwidth,clip]{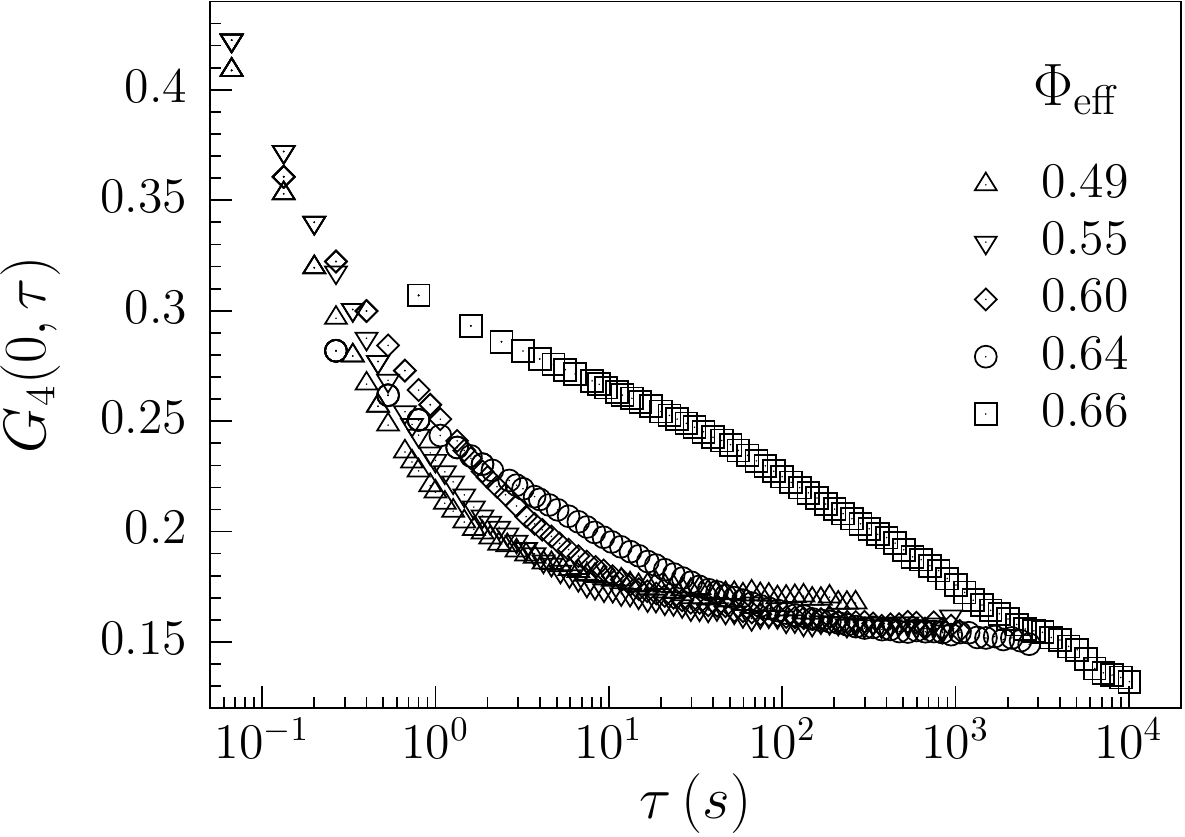}
\caption{Prefactor $G_{4}(0,\tau)$ in the microgel suspension as volume fraction increases. 
It decreases substantially with the lag time, and is found to depend
on volume fraction in the microgel suspension.}
\label{sigma2}
\end{figure}

Let us first focus on the direct measurement of
dynamical heterogeneities with a normalised 4-point correlation function
$G_4(r,\tau)/G_{4}(0,\tau)$~\cite{dasgupta_is_1991,JCP-Glotzer-2003}. 
Figure~\ref{g4_mes}-a shows $G_4(r,\tau^\star)/G_4(0,\tau^\star)$
in the microgel suspension with increasing volume fraction at
time $\tau^\star$, consistent with the peak value of $\chi_4$
(see inset of Figure~\ref{g4_mes}-b),
where dynamical heterogeneities are expected to be at their maximum. At low
volume fraction, $G_4(r,\tau^\star)/G_4(0,\tau^\star)$ exhibits an
exponential decay on a length scale $a= 0.31\,\mu$m 
which we will hereafter name 
``effective particle radius'' as it corresponds
approximately to the radius of the most abundant microgel $0.42\,\mu{\rm m}$
at $T=30\dg$C.
%
%
For the two highest volume fractions $\Phi_{\rm eff}=0.64$ and
$\Phi_{\rm eff} = 0.66$, the same exponential decay at small
$r$ is followed by a second decay at large $r$. This decay becomes
weaker with increasing volume fraction, directly suggesting the
increase of spatial correlations as the glass transition is
approached. 
The normalised spatial correlation then reaches the noise floor, 
which is around $10^{-3}$ for all volume fractions. 

The high volume fraction data in Figure~\ref{g4_mes}-a for
$G_4(r)$ were fit in the range $r = 3-10\,\micron$, corresponding to
the large $r$-regime, with the following family of functions, 
following Eq.~\eqref{eq:G4:Atau:exp:p}:
\begin{equation}
\label{functional1}
G_4({\rm r},\tau^\star)/G_4(0,\tau^\star) 
\propto K_{\Phi}  \left(\frac{a}{r}\right)^p 
\exp\left(-\frac{r}{\xi_4(\tau^\star)}\right)
\end{equation}
where the exponent $p$, the correlation length $\xi_4$, and 
the coefficient $K_{\Phi}$ are adjustable parameters~%
\footnote{%
To perform the fits,
the weight for each $G_4(r,\tau^\star)$ data point
was chosen as the number of pairs of ROIs
involved in the average over the whole sample 
($17\times 17$ ROIs), see Eq.~\eqref{G4-def-nonorm}.
}
. Values of $p$ close to zero were
found to provide the best adjustments and we set $p=0$. 
In order to check the robustness of the $\xi_4$ values
obtained from the $G_4$ measurements,
we added extra noise to the film images purposefully,
and it showed that the measurement of  $\xi_4$ is fairly robust against noise.
\footnote{Noise levels equal to 0.3, 1 and 3 
standard deviation of the original pixel intensity distribution were added.
At 0.3 and 1 standard deviations,
the resulting normalised $G_4(r,\tau^\star)$ profiles
(not shown) were unnoticeably altered
and the value of $\xi_4$ was correspondingly unaffected.
Only at a noise level of 3 standard deviations
was the first spatial decay of $G_4$ substantially stronger.}

The correlation length $\xi_4(\tau^\star)$,
shown in Figure \ref{g4_mes}-b and extracted from the fit,
increases significantly with volume fraction, 
and reaches a value
($29\pm 19$ microns) that exceeds
the size of the observed field
($13.6\times 13.6$ microns)
for the highest volume fraction investigated.

\section{Bad estimate of the spatial correlation length from the susceptibility $\chi_4$}

Let us estimate the growth of the correlation length 
that would be derived from $\chi_4$ if one were to use 
$2\pi\,\xi_4^2(\tau^\star) A =\chi_4(\tau^\star)$.
At low volume fraction $\Phi_{\rm eff}=0.49, 0.55$ and $0.6$, the peak value of $\chi_4$
is a constant. Since no correlations are expected, 
the correlation length is set equal 
to the effective particle radius, $\xi_4=a$. 
This also sets the value of the prefactor $A$. At higher
volume fraction, keeping the same value $A$ for the whole set of data
yields $\xi_4^{\Phi_{\rm eff}=0.64} = 1.16 
a$ 
and $\xi_4^{\Phi_{\rm eff}=0.66} = 1.41
a$.
The growth 
of this estimate of $\xi_4$ is widely
underestimated as compared to the direct measurement
(see Fig.~\ref{g4_mes}-b).


Given our experimental data for $G_4$, 
let us investigate the possible sources of error
and assess how reliably the
  growth of $\xi_4$ with volume fraction can be 
inferred from the growing peak in $\chi_4$. 
Since $\chi_4$ and $G_4$ are related through Eq.~\eqref{interpr-chi4},
let us investigate $G_4$ in more detail.
As
shown in 
Fig.~\ref{g4_mes}-a,
$G_4$ displays two spatial regimes for the highest volume fractions, with a crossover
radius $r_c$ corresponding to
approximately $5$ effective particle radii 
($a= 0.31\,\mu$m,
$r_c = 1.4\,\mu$m for $\Phi_{\rm eff}=0.66$ and $r_c =
1.7\,\mu$m for $\Phi_{\rm eff}=0.64$).
We now estimate both parts of the integral in
Eq.~\eqref{interpr-chi4}, $r<r_c$ and $r>r_c$,  as $\int_{0}^{r_c}2\pi r\exp(-r/a){\rm d}r $
and $\int_{r_c}^{\infty}2\pi r K_{\Phi}\,\exp(-r/\xi_4){\rm d}r $. 
The quantity $K_{\Phi}$ was found to depend on volume fraction, as 
$K_{\Phi=0.66}= 0.0105\pm 0.0016$
and 
$K_{\Phi=0.64}= 0.0063\pm 0.0015$.
For $\Phi_{\rm eff}=0.66$, we calculate that the $r<r_c$ part
of the integral in Eq.~\eqref{interpr-chi4} contributes about $1\%$ to the entire integral, 
while for $\Phi_{\rm eff}=0.64$, it represents more than $25\%$ of the entire integral. 
Even though this short distance contribution
which pollutes $\chi_4$ vanishes at large volume fraction,
it makes it risky to use $\chi_4$ for processing experimental results.
Indeed, in an experiment, the importance of this contribution 
cannot be estimated without measuring $G_4$ explicitly.

Another important source of error arises 
from the volume fraction dependence of the prefactor $A$
in Eq.~\eqref{chi4-sigma2B}.
Using Eq.~\eqref{functional1} with
$p=0$, Eq.~\eqref{chi4-sigma2B} writes:
\begin{equation}
2\pi\,\rho\,\xi_4^2(\tau^\star)
\sim\frac{\chi_4(\tau^\star)}{K_{\Phi}\,G_4(0,\tau^\star)}
\label{eq:xi4:squared:chi4bar}
\end{equation}
with $K_{\Phi}\,G_4(0,\tau^\star(\Phi))=A(\tau^\star)$.
The quantity $G_4(0,\tau)$, displayed in Figure
\ref{sigma2}, is found to decrease with the lag time 
$\tau$
and to depend on volume fraction. We find that
$
K_{0.64}
\,G_4(0,\tau^\star(0.64))=
0.0011$
at $\Phi_{\rm eff}=0.64$ and
$
K_{0.66}
\,G_4(0,\tau^\star(0.66))=
0.0020$
at $\Phi_{\rm eff}=0.66$. Since the prefactor
$A(\tau^\star)$ depends significantly on volume fraction, 
we can not obviously estimate the growth of $\xi_4$ 
with volume fraction, using $\chi_4(\tau^\star)$ alone.

\section*{Conclusion}

In conclusion, with the standard direct tool $G_4$, we measure the growth of
a dynamical correlation length $\xi_4$ 
from $1$ to $93 \pm 63$
effective particle radii with increasing volume fraction
in our soft microgel suspension.
This shows that the dynamics 
become
highly heterogeneous in space
when approaching the glass transition,
consistent with the broad theoretical picture \cite{bookheterog}.
Recently, various experimental studies have focused on the dynamical
susceptibility $\chi_4$ as a convenient indirect tool to quantify DHs,
mainly because the correlation function $G_4$ requires finer
measurements and data processing. Meanwhile, in various theoretical works,  
$\chi_4$ was understood to be not necessarily reliable 
\cite{heterogBerthier2,heterogDauchot2,PRL-Pastore-2011,Takeshi2013}.

Our results provide the first quantitative experimental evidence that
there is a significant error in estimating the dynamic correlation
length $\xi_4$ through $\chi_4$.
The origin of the failure of $\chi_4$ 
is related to the fact that
{\em (i)} it contains the correlations at the scale of the particle radius 
and {\em (ii)} the prefactor $A(\tau)$ in Eq. \eqref{chi4-sigma2B} varies with volume fraction.
It implies that
$\chi_4$ should {\em a priori never} be used as a
quantitative indicator of the increase of the spatial correlations,
before having first performed a direct measurement of $G_4$ to
estimate its validity. 
Nevertheless, as it is the variance of the time correlation function $Q_t(\tau)$,
the susceptibility $\chi_4$ is expected to remain an indicator 
of the lag time for which the dynamics is the most heterogeneous. 
In our experiments, the peak occurs approximately  at the suspension relaxation time
$\tau^\star$
defined by $Q(\tau^\star)=0.25$, see Fig.~\ref{functionQ}.

The two sources of error highlighted in this work should be kept in
mind when trying to relate the dynamic heterogeneities with other
features of the glass transition, such as {\it e.g.} the local
structure \cite{Royal-2015,Karmakar2014} or the soft modes of
vibration \cite{Chen2011-PRL,widmer-cooper_irreversible_2008}. 
They should be investigated in other glass forming materials 
and using other observable quantities.

\subsubsection*{Acknowledgements}
We thank E.~Bertin, L.~Cipelletti, O.~Dauchot, J.-B.~Fournier, 
F.~van~Wijland and L.~Wilson for fruitful discussions,
and A.~Callan-Jones for a critical reading of the manuscript. 
We acknowledge support
from PICS-CNRS {\em SofTher} (B.A. (P.I.), R.C. and A.A.).
\newline

\section{Appendix}

{Image pre-processing: The images are preprocessed in $4$ steps. 
We divide first the images by an average blank image taken with a water only sample 
to account for dust on the camera and spatial inhomogeneity in the illumination. 
The second step is an optional correction of the drift of the sample relative to the objective. 
We use the beads embedded in the suspension, 
which primarily are used for computing the volume fraction at $30\dg$C, to record the motion of the sample. 
Usually $2$ to $3$ beads are visible in the entire field of view of the CCD camera. 
We compute the average velocity of the centre of mass of the beads, 
smoothed over $1000$ frames using a temporal sliding window. 
A subset of the image, as large as possible to fit in the field of view while being advected 
is extracted by advecting the subframe at the drift speed, using a bilinear pixel interpolation 
to extract pixel values. Further preprocessing consists in applying $2$ filters. 
A rolling ball background subtraction (radius $60$ px) to eliminate heterogeneities in illumination 
coming from out-of-focus beads, and a Savitzky and Golay filter 
(local polynomial interpolation, width$=5$ px, order $= 4$) 
is applied to reduce the effect of the camera shot noise. }





\bibliography{biblio-all_complet,biblio-all}
\bibliographystyle{rsc} 


\end{document}